\documentclass[prd, twocolumn,10pt]{revtex4-1}
\usepackage{epsfig}
\usepackage{amsmath}
\usepackage{amsfonts}
\usepackage{graphicx}
\usepackage{dsfont}
\usepackage[german, english]{babel}
\usepackage{amssymb}
\usepackage{ifthen}
\usepackage{ulem}
\usepackage{color}
\usepackage{float}
\usepackage{physics}

\newcommand{\figref}[1]{Figure \ref{#1}}
\renewcommand{\eqref}[1]{Eq. (\ref{#1})}

\usepackage{caption}
\RequirePackage[colorlinks=true,allcolors=blue,
unicode=true,hypertexnames=false]{hyperref}
\hypersetup{pdfstartview={XYZ null null 1.25}}

\usepackage{float} 
\usepackage{caption} 
\usepackage{subcaption} 

\usepackage{xcolor}
\usepackage{soul}

\begin{document}

\title{Reply to ``Comment on `Surface Pair-Density-Wave Superconducting and Superfluid States' ''}

 \author{Mats Barkman}
 \affiliation{Department of Physics, Royal Institute of Technology, SE-106 91 Stockholm, Sweden}
 
 \author{Andrea Benfenati} 
 \affiliation{Department of Physics, Royal Institute of Technology, SE-106 91 Stockholm, Sweden}

 \author{Albert Samoilenka}
 \affiliation{Department of Physics, Royal Institute of Technology, SE-106 91 Stockholm, Sweden}

 \author{Egor Babaev}
 \affiliation{Department of Physics, Royal Institute of Technology, SE-106 91 Stockholm, Sweden}

\maketitle

By using an approximate quasiclassical treatment, the Comment asserts that surface pair-density-wave (PDW) states are not supported by microscopic theory.
We demonstrate below that the claim is incorrect by both pointing out that fully microscopic demonstrations of surface PDW states already exist in tight binding models \cite{samoilenka2020pair}, and providing a fully microscopic solution for the continuum case, considered in the Comment, but without relying on 
the approximations made there.

Three questions are raised in the Comment concerning the Ginzburg-Landau (GL) approach.
Point (i) highlights that we use $H_{FFLO}(T)\approx H_0(T)$ in \cite{barkman2018surface}, which is justified close to the tri-critical point. Indeed, we use that approximation because it is that regime where the GL theory itself can be used justifiably. 
Nonetheless, using non-approximated coefficients, as we do in \mbox{\cite{samoilenka2020pair}} Eqs.(2-4), obviously does not make any difference for the question of  existence of surface PDW states.
Point (ii) addresses the use of truncated gradient expansion. 
Evidently the GL part of our work, by construction, applies in the vicinity of the tri-critical point, where the order parameter varies on macroscopic lengthscales. Thus,   a truncated gradient expansion, is justified   for the description of long-wavelength physics in that regime.
{Our follow-up \cite{samoilenka2020pair}    demonstrated the effect from a completely microscopic point of view, that does not rely on any expansion.}
Point (iii) asserts that the GL theory  we use was obtained with bulk Green's functions.
This was explicitly pointed out in  \cite{samoilenka2019boundary,samoilenka2020pair} along with the 
ambiguities that arise when employing the phenomenological treatment of boundary conditions. 
{That is why in {\mbox{\cite{samoilenka2020pair}}} we obtained  surface states in  fully microscopic theory.}
The discussion on boundary conditions in {\mbox{\cite{samoilenka2019boundary}}} was separated from the published version {\mbox{\cite{samoilenka2019Tc2}}} and is now available with the full solution with the microscopically derived GL boundary conditions {\mbox{\cite{samoilenka2020microscopic}}}.

 There already exists fully microscopic demonstration of the surface states 
 {\mbox{\cite{samoilenka2020pair}}}.
 By contrast the Comment is using a quasiclassical approximation that, as we already pointed out in {\mbox{\cite{samoilenka2019Tc2}}}, if straightforwardly applied,     in general misses the surface states with elevated critical temperatures.
The surface states should be obtainable within the quasiclassical approach, but that requires a more accurate  treatment of the boundary conditions that fully reflects the microscopic physics of the interface, analogous to the discussion in {\mbox{\cite{samoilenka2020microscopic}}}.

Furthermore, we demonstrated, at a fully microscopic level,  enhanced boundary critical temperature for 
a superconductor  in the absence of Zeeman splitting in {\mbox{\cite{samoilenka2020pair,samoilenka2019Tc2}}}, showing that the effect does not rely on FFLO physics.
The Comment cites \cite{samoilenka2020pair,samoilenka2019Tc2}
and correctly observes our   conclusion reached therein that FFLO physics is not the primary mechanism for the enhancement effect.
However the interpretation in terms of the critical temperature of a surface atomic layer in the footnote of the Comment is incorrect. 
In particular in {\mbox{\cite{samoilenka2019Tc2}}}, the effect is demonstrated in continuum theory, and the boundary states arise from properly applying vanishing boundary conditions to the quasi-particle wavefunctions $u_\uparrow (0) = v_\downarrow (0) = 0$.

\begin{figure}
    \centering
    \includegraphics[width=0.49\textwidth]{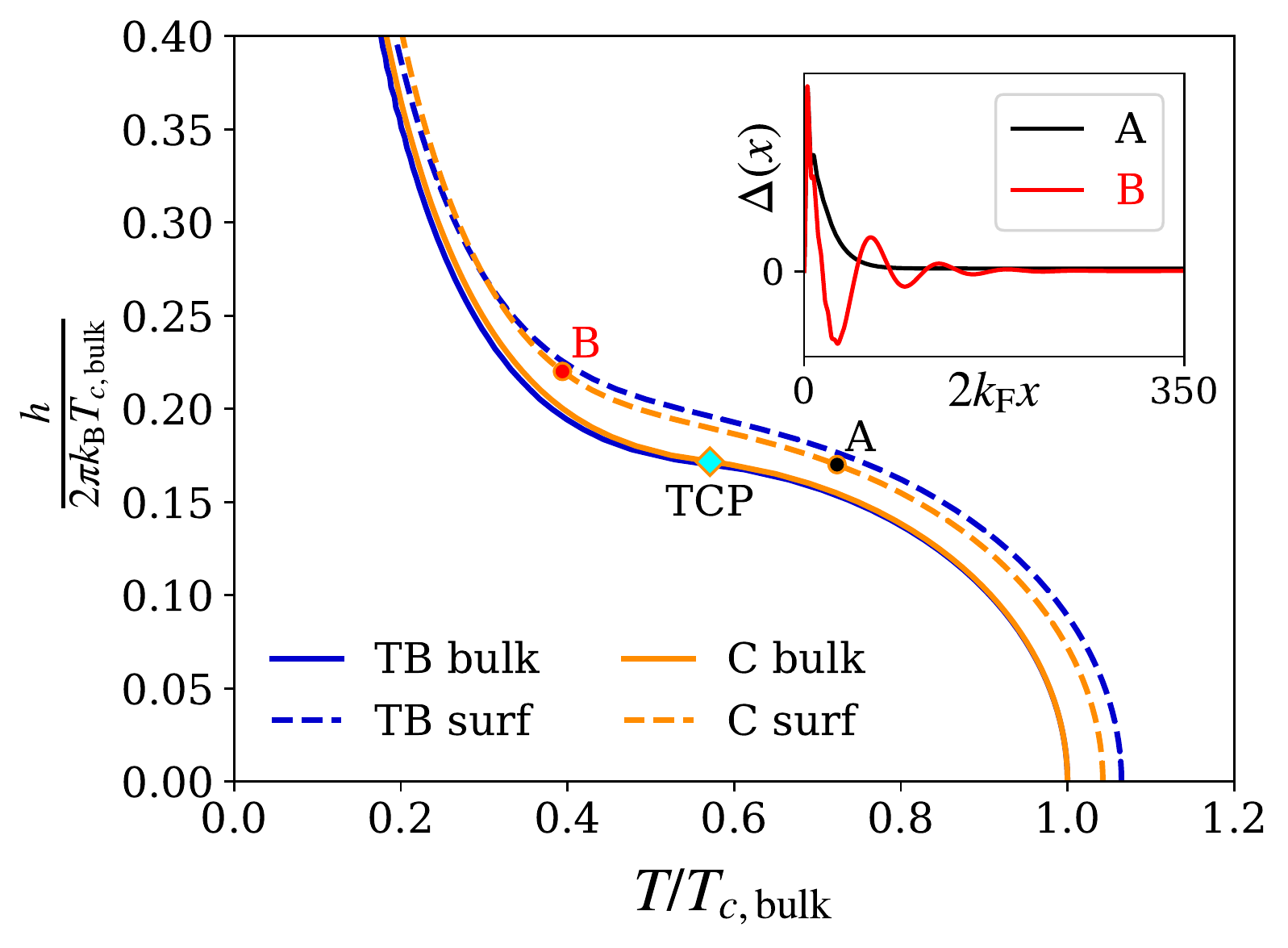}
    \caption{Phase diagram obtained by solving the microscopic Bogoluibov-de Gennes equations, demonstrating enhanced critical temperature for surface states. $T$ and $h$ denote temperature and Zeeman splitting energy. The enhancement is demonstrated both for a tight-binding (TB) and a continuum (C) model. The tri-critical point in the continuum model is denoted TCP. The inset shows the order parameter profile at points A and B.} \label{fig: phase diagram}
\end{figure}

Lastly we demonstrate the existence of surface states by solving the microscopic Bogoliubov-de Gennes (BdG) equations for an imbalanced superconductor, both in a continuum almost free fermion model and a tight binding lattice model \footnote{Note that tight-binding simulations of this kind, along with demonstration of the phase diagram already have been presented in {\mbox{\cite{samoilenka2020pair}}} for different parameters}.
The BdG equations for the quasi-particle wavefunctions read
\begin{equation}
    \begin{pmatrix}
    H_+ & \Delta (x) \\
    \Delta^* (x) & -H_-
    \end{pmatrix}
    \begin{pmatrix}
    u_\uparrow (x) \\ v_\downarrow (x)
    \end{pmatrix}
    = 
    E
    \begin{pmatrix}
    u_\uparrow (x) \\ v_\downarrow (x)
    \end{pmatrix}
\end{equation}
where $H_\pm = \epsilon - (\mu \pm h)$, with $\epsilon = - \frac{\hbar^2 \nabla^2}{2m}$ in the continuum model, and $\epsilon = -t ( \delta_{i,i+1} + \delta_{i,i-1} )$ in the tight-binding model. The gap parameter is determined self-consistently through $\Delta (x) = V \expval{\Psi_\uparrow (x) \Psi_\downarrow (x)}$
\footnote{ The critical temperature and gap parameter profiles at the transitions are found by solving the corresponding linearized gap equation. For more details on the linearized gap equation, see {\mbox{\cite{samoilenka2019Tc2}}}. For the continuum model we set $V=0.3 \pi \hbar v_F$ and for the tight-binding model we set $V = \mu = t$. The bulk critical temperatures in the absence of Zeeman splitting are $k_{\rm{B}}T_{c,\rm{bulk}} \simeq 0.159 \mu$ in the continuum model, and $k_{\rm{B}}T_{c,\rm{bulk}} \simeq 0.0147 t$ in the tight-binding model. These temperatures are used to rescale \figref{fig: phase diagram}.}.
The demonstration of surface states in both of these systems, as in  {\mbox{\figref{fig: phase diagram}}}, directly disproves the claim in the Comment that the bulk instability is the most favorable one.

\begin{acknowledgments}
 
\end{acknowledgments}
The work was supported by the Swedish Research Council Grants No. 642-2013-7837, 2016-06122, 2018-03659.  

\bibliography{references}

\end{document}